\documentclass[a4paper]{article}

\usepackage[T1]{fontenc} 
\usepackage{float}       
\usepackage{helvet}      
\usepackage{mathptmx}    
\usepackage{rsc}         
\usepackage{setspace}    
\AtBeginDocument{\doublespacing}

\newfloat{chart}{htbp}{loc}
\newfloat{graph}{htbp}{loh}
\newfloat{scheme}{htbp}{los}

\usepackage{graphics,pstricks,amsmath}
\usepackage{graphicx}
\usepackage{epsfig}

\author{%
  Piotr S. \.Zuchowski\thanks{%
    Department of Chemistry, Durham University%
    E-mail: \texttt{piotr.zuchowski@dur.ac.uk}
  }%
\ and 
  Jeremy M. Hutson\thanks{%
    Department of Chemistry, Durham University%
    E-mail: \texttt{j.m.hutson@dur.ac.uk}
  }%
}

\title{Cold collisions of N ($^4S$) atoms and NH ($^3\Sigma$) molecules in magnetic fields}

\begin{document}

\maketitle

\begin{abstract}
We calculate the interaction potential between N atoms and NH
molecules and use it to investigate cold and ultracold
collisions important for sympathetic cooling. The ratio of
elastic to inelastic cross sections is large over a wide range
of collision energy and magnetic field for most isotopic
combinations, so that sympathetic cooling of NH molecules by N
atoms is a good prospect. However, there are important effects
due to a p-wave shape resonance that may inhibit cooling in
some cases. We show that scaling the reduced mass used in the
collision is approximately equivalent to scaling the
interaction potential. We then explore the dependence of the
scattering properties on the reduced mass and explain the
resonant effects observed using angular-momentum-insensitive
quantum defect theory.
\end{abstract}

\section{Introduction}
\label{sec:intro}

At temperatures below about 1 mK, atoms and molecules enter a
fully quantal regime where their de Broglie wavelength is large
compared to molecular dimensions. In this regime, collision
cross sections and reaction rates are dominated by long-range
forces \cite{Wigner:1948, Balakrishnan:threshold:1997} and
resonance phenomena \cite{Forrey:1998}. It is likely to be
possible to {\em control} reaction rates by tuning scattering
resonances with applied electric and magnetic fields
\cite{Krems:IRPC:2005, Gonzalez-Martinez:2007,
Hutson:HeO2:2009}. At even lower temperatures, below about 1
$\mu$K, trapped atoms and molecules form {\em quantum gases}
such as Bose-Einstein condensates and Fermi-degenerate gases,
in which every molecule occupies the lowest allowed
translational state in the trap. The quantum gas regime offers
additional possibilities for a new form of quantum control, in
which chemical transformations are carried out coherently on
entire samples of ultracold atoms and molecules.

There have been enormous advances towards these goals in the
last few years. In particular, it is now possible to produce
ultracold alkali metal dimers in their rovibronic ground states
from ultracold atoms both by photoassociation \cite{Sage:2005,
Deiglmayr:2008, Viteau:2008} and by magnetoassociation followed
by stimulated Raman adiabatic passage (STIRAP)
\cite{Ni:KRb:2008, Danzl:ground:2010}. However, for an alkali
metal dimer even the ground rovibronic state has nuclear spin
hyperfine structure \cite{Aldegunde:polar:2008,
Aldegunde:nonpolar:2009}, and the resulting splittings are
comparable to the kinetic energies involved in ultracold
collisions. For the case of $^{40}$K$^{87}$Rb, microwave
transitions have been used to transfer the ground-state
molecules selectively between different hyperfine and Zeeman
levels \cite{Aldegunde:spectra:2009, Ni:KRb:2008}. In a very
recent development, Ospelkaus {\em et al.}\
\cite{Ospelkaus:react:2010} have studied reactive collisions of
such state-selected molecules, both with one another and with
ultracold K and Rb atoms. They observed remarkable selectivity
of the resulting reactions, in which flipping the spin of a
single nucleus could cause dramatic changes in the outcome of a
collision \cite{Hutson:KRb-perspective:2010}.

Methods that form molecules from ultracold atoms can be applied
only in cases where the atoms themselves can be cooled. In
practice this restricts them to the alkali metals, the alkaline
earths, and a few other elements. These species have a fairly
limited chemistry. In order to cool a wider class of molecules,
including polyatomic ones, a number of {\em direct} cooling
methods \cite{Bethlem:1999, Weinstein:CaH:1998, Egorov:2002,
Elioff:2003, Junglen:2004} have been established over the last
decade. Among these methods, {\em buffer-gas cooling} is based
on the particularly simple idea of cooling molecules by elastic
collisions with cold He gas. If the molecules are paramagnetic
and in low-field-seeking states, they can be confined in a
magnetic trap. The temperatures which can be achieved in buffer
gas cooling method are limited by the vapour pressure of the
buffer gas (ca.\ 400 mK for $^3$He), but the method is
particularly valuable for two reasons: (i) it is very general
and can in principle be applied to any paramagnetic species,
provided that a detection scheme is available; (ii) it can
produce large numbers and high densities of cold molecules.
Buffer-gas cooling has been reported for a variety of molecules
including CaH \cite{Weinstein:CaH:1998}, CaF
\cite{Maussang:2005}, NH \cite{Egorov:2004, Campbell:2007} ,
CrH and MnH \cite{Stoll:2008} and also for a number of
paramagnetic atoms \cite{Kim:Eu:1997, Weinstein:Cr:2002,
Hancox:Mo:2005}. Buffer-gas cooling followed by evaporative
cooling has recently been used to achieve Bose-Einstein
condensation with no laser cooling for metastable helium
\cite{Doret:2009}.

The direct methods established so far are limited to
temperatures of 10 to 100 mK and above. To cool the molecules
further, to the $\mu$K regime, second-stage cooling methods
must be developed. The most promising and conceptually the
simplest method is {\em sympathetic cooling}, in which the
molecules are cooled by collisions with an atomic gas that can
itself be cooled to the ultracold regime, such as an alkali
metal. The most robust trapping methods for molecules work for
low-field-seeking states, which are never the lowest possible
state in an applied field. Inelastic collisions can therefore
occur, and either heat the trapped system or eject the
molecules from the trap. Sympathetic cooling can thus be
successful only if elastic collisions dominate inelastic ones,
and it is usually stated that the ratio of elastic to inelastic
cross sections must be 100 or more. Sympathetic cooling was
initially developed as a cooling method for trapped ions
\cite{Larson:1986}. More recently it has been used to achieve
sub-Kelvin temperatures for polyatomic ions
\cite{Ostendorf:2006} and has also been used to produce
ultracold neutral atoms with scattering properties that are not
suitable for evaporative cooling, such as $^{41}$K
\cite{Modugno:2001}.

Sympathetic cooling for molecules has not yet been achieved,
but several proposals have been explored. It was initially
proposed for NH molecules colliding with Rb atoms
\cite{Soldan:2004} and studied in more details by Lara {\em et
al.}\ \cite{Lara:PRL:2006, Lara:PRA:2007} for OH colliding with
Rb. Both OH and NH molecules interact very strongly with Rb and
the anisotropy of the interaction potential is large compared
to the molecular rotational constant. The large anisotropy
implies large couplings between channels with different $n$
(monomer rotation angular momentum) and $L$ (end-over-end
angular momentum) quantum numbers, and Lara {\em et al.}\
showed that this resulted in large inelastic cross-sections in
the ultracold regime. The remedies they suggested to improve
sympathetic cooling and decrease inelastic cross sections were:
(i) to use light atoms as coolants, in order to increase the
heights of centrifugal barriers and suppress inelastic
channels; (ii) to find atom-molecule system with much smaller
anisotropy in the interaction potential.

Sold\'an \textit{et al.}\ \cite{Soldan:2009} considered the
possibility of reducing the anisotropy by using alkaline-earth
atoms (Ae) as collision partners for NH molecules. They showed
that the neutral states of Ae--NH systems are coupled to
ion-pair states Ae$^+$NH$^-$, with crossings between the
neutral and ion-pair surfaces at linear geometries. For Sr and
Ca atoms the crossings occurs at energies below the
atom-molecule threshold, so will be accessible in low-energy
collisions. However, for Be-NH and Mg-NH the crossings occurs
at energies more than 1000 cm$^{-1}$ above the atom-molecule
threshold. In these systems, the ion-pair state is likely to be
inaccessible, so it is reasonable to carry out collision
calculations on a single covalent surface. In addition, the
potential energy surface for Mg--NH turned out to be only
weakly anisotropic. Wallis and Hutson \cite{Wallis:MgNH:2009}
carried out quantum scattering calculations of spin relaxation
collisions (in magnetic fields) and showed that sympathetic
cooling of NH by collisions with Mg atoms should be achievable
if the molecules can be precooled to about 10~mK.

Sympathetic cooling has also been considered for NH$_3$ and
ND$_3$. In this case the molecules are initially slowed in a
Stark decelerator \cite{Bethlem:tvar:2002}. \.Zuchowski {\em et
al.}\ \cite{Zuchowski:NH3:2008} surveyed the interaction
potentials for NH$_3$ interacting with alkali-metal and
alkaline-earth atoms. \.Zuchowski and Hutson
\cite{Zuchowski:2009} then carried out quantum scattering
calculations on collisions of ND$_3$ with Rb atoms and showed
that molecules that are initially in the upper component of the
ammonia inversion doublet are likely to undergo fast
collisional relaxation to the ground state, and that this is
likely to prevent sympathetic cooling of molecules trapped in
low-field-seeking states in an electrostatic trap
\cite{Bethlem:trap:2000}. However, there is a good prospect for
sympathetic cooling of ammonia molecules in high-field-seeking
states, even with magnetically trapped atoms, because the terms
in the hamiltonian that might cause spin-changing collisions of
the Rb atoms are very small. High-field-seeking states of
ND$_3$ can be confined in an alternating current trap
\cite{vanVeldhoven:2005}.

Recently, Hummon {\em et al.}\ \cite{Hummon:2009} demonstrated
buffer-gas cooling and trapping of N ($^4$S) atoms and
simultaneous co-trapping of NH molecules. Subsequent work
\cite{Tscherbul:NN:2010} has demonstrated N atom densities
around $5\times10^{12}$ cm$^{-3}$ and lifetimes around 10~s.
This offers the possibility of cooling the atoms further with
atomic evaporative cooling, which has already been achieved for
metastable helium and Cr atoms \cite{Nguyen:2005,
Nguyen:Cr:2007}.

A gas of N atoms is potentially an excellent coolant for a
sympathetic cooling experiment. The N atom has a very small
polarizability compared to Group I and Group II elements and
this results in low $C_6$ coefficients and small anisotropies
of the interaction potentials with molecules. The N atom also
has a relatively low mass, which results in higher centrifugal
barriers and stronger suppression of inelasticity for particles
scattered with $L>0$.

This paper presents theoretical studies of cold and ultracold
collisions of N atoms with NH molecules, in order to
investigate the possibility of sympathetic cooling of NH by
atomic nitrogen. Since the cross sections depend strongly on
the reduced mass of the collision system, we consider four
isotopic combinations of N--NH systems, with each of the two N
atoms being either $^{14}$N or $^{15}$N. We assume that both N
and NH are initially in their magnetically trappable
spin-stretched states, with the maximum possible values of the
electron spin projection numbers. For such states only spin
relaxation (and not spin exchange) can occur and only the
sextet interaction potential contributes. We report
calculations of the sextet potential for N--NH and explore the
behaviour of cross sections as a function of collision energy
and magnetic field. We discuss the sensitivity of the
scattering results with respect to uncertainties in the
interaction potential. Finally, we analyze the behaviour of the
shape resonances in terms of angular-momentum-insensitive
quantum-defect theory (AQDT) \cite{Gao:2001}.

\section{N--NH potential}

The total spin of the N($^4$S) + NH($^3\Sigma^-$) system can be
$\frac{5}{2}$, $\frac{3}{2}$ or $\frac{1}{2}$. The chemical
reaction N~+~NH~$\to$~N$_2$~+~H, which occurs principally on
the doublet surface, has been studied in detail by Varandas and
coworkers \cite{Poveda:NNH:2003, Caridade:NNH:2005,
Caridade:NNH:2007} and by Francombe and Nyman
\cite{Francombe:2007}. It was shown that the doublet N--NH
system forms an N$_2$H complex without a potential barrier
along the minimum energy path. A very small barrier exists
between the N$_2$H complex and N$_2$ + H products and overall
the reaction of forming N$_2$+H yields 6.33 eV of energy. To
our knowledge, no studies of quartet or sextet states of N--NH
have been published, though we are aware of work in progress by
Tscherbul and coworkers \cite{Tscherbul:private:2010}.

To obtain the sextet interaction potential we applied the
recently developed explicitly correlated, unrestricted
coupled-cluster method with single, double and noniterative
triple excitations [UCCSD(T)] \cite{Knizia:2008, Knizia:2009,
Adler:2007}. We used the aug-cc-pVTZ basis set of Peterson {\em
et al.}\ \cite{Peterson:2008}, which is designed specifically
for use with explicitly correlated calculations. The results
from the explicitly correlated (F12) calculation are compared
with those from UCCSD(T) calculations with uncorrelated basis
sets in Table \ref{basconv}: it may be seen that the explicitly
correlated approach dramatically reduces the error caused by
using unsaturated basis sets. A fixed NH bond length of 1.0367
\AA\ was used in all the calculations.

\begin{table}
\caption{Basis-set dependence of the N--NH interaction energy
at the global minimum for F12 calculations. The complete basis
set (CBS) extrapolation was obtained with the correlation
energy functional $E(X)=A + B X^{-3}$ \cite{Helgaker:1997:CBS}
where $X$ is the maximum angular momentum of electronic basis
set.}
 \label{basconv}
   \centering
   \begin{tabular}{ll} \hline
    basis set        & $E_{\rm int}$ (cm$^{-1}$)      \\ \hline
    aug-cc-pVTZ      &  $-79.18$              \\
    aug-cc-pVQZ      &  $-86.03$              \\
    aug-cc-pV5Z      &  $-88.51$              \\
    aug-cc-pV6Z      &  $-89.33$              \\
    CBS              &  $-90.47$              \\
    F12 /aug-cc-pVTZ &  $-89.10$              \\\hline
   \end{tabular}
\end{table}

The potential energy surface was obtained by carrying out
explicitly correlated UCCSD(T) calculations on a grid in Jacobi
coordinates ($R_i,\theta_j)$, where $R$ is the intermolecular
distance measured to the NH center of mass and $\theta$ is the
angle between the NH bond vector and the vector from the NH
center of mass to the N atom. The radial grid $R_i$ was from
2.5 to 10 \AA\ in 0.25 \AA\ steps and the angular grid
$\theta_j$ was a set of 11 Gauss-Lobatto quadrature points,
which include the two linear geometries. All interaction
energies were corrected for basis-set superposition error using
the counterpoise procedure \cite{Boys:1970}.

Radial interpolation is carried out using the reproducing
kernel Hilbert space (RKHS) method \cite{Ho:1996, Soldan:2000}
to evaluate $V(R,\theta_j)$ for arbitrary $R$ and given
$\theta_j$. At each distance $R$, the potential is expanded in
Legendre polynomials $P_\lambda(\cos \theta)$ for $\lambda$ up
to 8,
\begin{equation} V(R,\theta)= \sum_\lambda V_\lambda(R)
P_\lambda(\cos\theta). \end{equation} The coefficients
$V_\lambda(R)$ are obtained by integrating the {\em ab initio}
potential using Gauss-Lobatto quadrature \cite{Lara:PRA:2007}.

To provide an improved description of the long-range
interaction, we impose an analytical representation on the
long-range part of the components of the projected potential,
\begin{equation}
V^{\rm lr}_\lambda (R) = - \sum_{n=6}^{8} \sum_{\lambda=0}^{3} C_{n,\lambda} R^{-n}.
\end{equation}
The Van der Waals coefficients are given in Table \ref{vdwcoef}
and were calculated with the restricted open-shell coupled
Kohn-Sham method \cite{Zuchowski:SAPT-DFT:2008} with
asymptotically corrected \cite{Tozer:1998} PBE0 functional
\cite{Adamo:1999}. We connect the long-range function smoothly
to the supermolecular potential using the switching function
\cite{Janssen:2009}
\begin{eqnarray}
f(R) = \frac{1}{2}  + \frac{1}{4} \sin \frac{\pi x}{2} \big(3-\sin^2 \frac{\pi x}{2} \big),
\end{eqnarray}
where $x=\frac{R-b+ R-a}{b-a}$ with $a=7$ \AA\ and $b=11$ \AA.
$f(R)=0$ for $R<7$ \AA\ and $f(R) =1$ for $R>11$ \AA.

\begin{table}
\caption{Van der Waals coefficients for N--NH ($E_{\rm h}
a_0^n$) from density-functional calculations. }
 \label{vdwcoef}
   \centering
   \begin{tabular}{ll}\hline
    $n,\lambda$   &     $C_{n,\lambda}$   \\ \hline
    $6,0$         &         33.50         \\
    $6,2$         &         11.44         \\
    $7,1$         &         60.87         \\
    $7,3$         &         55.28         \\
    $8,0$         &         717.14        \\
    $8,2$         &         988.87        \\ \hline
   \end{tabular}
\end{table}

\begin{figure}
\includegraphics[width=\linewidth]{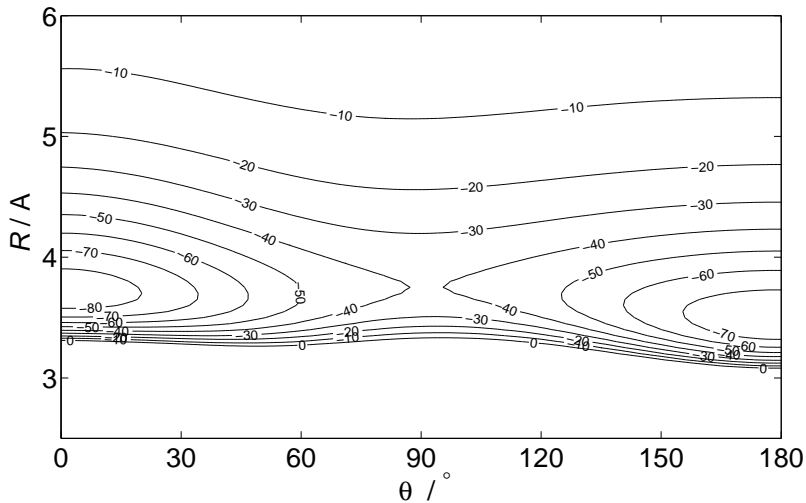}
\caption{ The {\it ab initio} interaction potential of N--NH.
 Contours are labeled in cm$^{-1}$. The angle $\theta=0$
corresponds to the N--N-H geometry.  }
\label{potential}
\end{figure}

The potential energy surface for N--NH is shown in Fig. \ref{potential}.
 It has two minima of comparable depths at
linear geometries: 89.1 cm$^{-1}$ at N--NH and 76.4 cm$^{-1}$
at N--HN. The two minima are separated by a saddle point near
the T-shaped geometry. The anisotropy of the potential near the
Van der Waals minimum is about 40 cm$^{-1}$, and the dominant
contribution to the anisotropy arises from $V_2(R)$.

\begin{table}
\caption{Characteristic points on the N--NH potential energy
surface. Energies are given in cm$^{-1}$, $R$ in \AA.}
 \label{geoms}
   \centering
   \begin{tabular}{llll}
 \hline
                     &  Global minimum  & Saddle point & Secondary minimum \\ \hline
    $R,\theta$       &  3.70,0  & 3.76, 92$^\circ$ & 3.49, 180$^\circ$  \\
                     &  -89.1 & -39.2          & -76.4            \\ \hline
   \end{tabular}
\end{table}

\subsubsection{Interaction potential uncertainty}

An important problem in electronic structure theory is the
estimation of error bounds for calculated interaction energies.
Since scattering calculations at very low energies depend
strongly on the details of the interaction, in this section we
discuss the uncertainty of the N--NH interaction potential
obtained here.

The largest contributions to the uncertainty of the interaction
potential arise from the approximate treatment of electronic
correlation and the incompleteness of the electronic basis set.
We expect that the effect of the neglecting vibrations of the
NH molecule is much less important, as are relativistic and
nonadiabatic effects.

First we need to estimate how well the UCCSD(T) method works
for sextet N--NH. To explore this, we performed 7-electron full
configuration-interaction (FCI) calculations of the interaction
energy. We included all electrons arising from the H atom and
the $2p$ electrons of N atoms. With FCI it was possible to use
only a very small basis set (6-31G, augmented with $spd$
midbond functions with an exponent~0.4). For this small basis
set, we compared the contribution to the correlation part of
the interaction energy (i.e., the intramonomer correlation and
the dispersion energy) with the UCCSD(T) results for several
linear geometries N--NH and N--HN . The FCI correlation energy
to be larger than the UCCSD(T) correlation energy by 1 to
1.5\%, for a wide range of $R$ and at both linear geometries.
To a good approximation we expect the {\em ratio} $E_{\rm
corr}^{\rm FCI}/E_{\rm corr}^{\rm UCCSD(T)}$ to be constant in
different basis sets. This suggests that the global minimum
energy obtained with the coupled-cluster method is
underestimated by ca.\ 1.5 cm$^{-1}$.

The basis set convergence pattern shown in Table \ref{basconv}
yields a complete basis-set limit of the global minimum depth
of 90.47 cm$^{-1}$. This is 1.37 cm$^{-1}$ more than in the
method used for the complete surface here. We also performed
test calculations including additional core-valence basis
functions that are absent in the basis set used for the
complete surface potential. The interaction energy at the
global minimum obtained with aug-cc-pCVTZ is approximately
1~cm$^{-1}${\textit smaller} than for basis sets with no
core-valence functions.

In summary we can set the error bounds on the interaction
potential at the global minimum between $-1$ and +3 cm$^{-1}$,
which is approximately between $-1$\% and +3\%.

\section{N--NH scattering calculations}
\label{sec:theory} The Hamiltonian of the NH molecule may be
written
\begin{eqnarray}
 H_{\rm NH} &=& b_{\rm NH} \hat N^2 + \gamma \hat N \cdot \hat S \nonumber\\ &&+
 \left[ \frac{96 \pi}{45} \right]^\frac{1}{2} \lambda_{\rm SS} \sum_q
(-1)^q Y_{2,-q} (\hat r) \left[ S \otimes S \right]^{(2)}_q.
\label{ham-mono}
\end{eqnarray}
The three terms are, respectively, the rigid rotor Hamiltonian,
the spin-rotation interaction and the intramonomer spin-spin
interaction. The numerical values of the constants used in the
present work are $b_{\rm NH}=16.343 $ cm$^{-1}$
\cite{Brazier:1986}, $\gamma=-0.055$ cm$^{-1}$ and
$\lambda_{\rm SS}=0.92$ cm$^{-1}$ \cite{Mizushima}. The NH
molecule is assumed to be in its ground vibrational state.

The Hamiltonian of the N--NH collision system in a magnetic
field may be written
\begin{equation}
H= -\frac{\hbar^2}{2 \mu R } \frac{d^2}{dR^2} R + \frac{\hat L^2}{2\mu R^2}
+ H_{\rm NH} + H_{\rm Z}
+ V_{\rm SS} + V_{\rm int}(R,\theta).
\label{ham-nnh}
\end{equation}
Here $\hat L^2$ is the operator for the end-over-end angular
momentum of N and NH about one another, $H_{Z}$ represents the
Zeeman interaction of N and NH with the magnetic field, $V_{\rm
SS}$ is the (anisotropic) intermolecular spin-spin interaction,
and $ V_{\rm int}(R,\theta)$ is the intermolecular potential.

The convention for quantum numbers in this paper is as follows:
$L$ and $M_L$ denote the end-over-end angular momentum and its
projection onto the space-fixed $Z$ axis defined by the
magnetic field. Monomer quantum numbers are indicated with
lower-case letters to avoid confusion with those of the
collision system as a whole. The spins and spin projections of
the N and NH molecules are denoted by $s_{\rm A} $, $s_{\rm B}
$ and $m_{s {\rm A}}$, and $m_{s {\rm B}}$, respectively. The
rotational quantum number of the NH molecule and its projection
are denoted $n_{\rm B}$ and $m_{n \rm B}$  The projection of
the total angular momentum,
\begin{equation}
M_{\rm tot} = M_L + m_{n \rm B} + m_{s {\rm B}} +  m_{s {\rm A}},
\end{equation}
is rigorously conserved in a collision, but the total angular
momentum itself is not, except at zero field. It is convenient
to carry out scattering calculations is a fully uncoupled basis
set, $|s_{\rm A} m_{s {\rm A}}\rangle |s_{\rm B} m_{s {\rm
B}}\rangle |n_{\rm B} m_{ n{\rm B} }\rangle |L M_L \rangle$. We
have written a plug-in routine for the MOLSCAT scattering
program \cite{MOLSCAT:v15}, implementing all the matrix
elements required for scattering calculations in this basis
set.

set.

The total spin $S$ of a system made up of an open-shell atom
and an open-shell molecule can take values between $|s_{\rm A}
- s_{\rm B}|$ and $s_{\rm A}+s_{\rm B}$. For N--NH the allowed
values are $S=\frac{1}{2}$, $\frac{3}{2}$ and $\frac{5}{2}$,
corresponding to doublet, quartet and sextet, respectively. The
interaction potential $ V_{\rm int} (R,\theta)$ may be written
in terms of projection operators,
\begin{equation}
  V_{\rm int} (R,\theta) = \sum_{S=-| s_{\rm A} + s_{\rm B} |  }^{s_{\rm A} + s_{\rm B} }
 |S \rangle V_S(R,\theta) \langle S |
\end{equation}
and the general matrix element of  $ V_{\rm int} (R,\theta)$ in
our basis set is
\begin{eqnarray}
\langle s_{\rm A} m_{s{\rm A}} s_{\rm B} m_{s{\rm B}} n_{\rm B}  m_{n {\rm B}} L M_L |V_{\rm int} (R,\theta)|
 s_{\rm A} m'_{s_{\rm A}} s_{\rm B} m'_{s{\rm B}} n'_{\rm B}  m'_{n {\rm B}} L' M'_L \rangle =
\nonumber \\
\sum_S
(-1)^{2s_{\rm A} + 2 s_{\rm B} -  m_{s{\rm A}} -  m_{s{\rm B}} - M_L  }  (2S+1)
 \langle n_{\rm B}  m_{n{\rm B}} L M_L  |  V_S (R,\theta) | n'_{\rm B}  m'_{n{\rm B}} L' M'_L  \rangle \nonumber \\
\left( \begin{array}{ccc}  s_{\rm A}& s_{\rm B}& S \\ m_{s{\rm A}} & m_{s{\rm B}} & -m_{s{\rm A}}-m_{s{\rm B}} \end{array}
   \right)
\left( \begin{array}{ccc}  s_{\rm A}& s_{\rm B}& S \\ m'_{s{\rm A}}&  m'_{s{\rm B}}&  -m'_{s{\rm A}}-m'_{s{\rm B}} \end{array}
   \right).
\label{intpot}
\end{eqnarray}
The three interaction potentials $V_S(R,\theta)$ differ only by
short-range Pauli exchange terms. They have the same long-range
coefficients, so become degenerate once the N atom and NH
molecule are far enough apart that their valence shells do not
overlap. The doublet surface has a potential well several
hundred times deeper than the Van der Waals sextet state, so
that full quantum calculations including the doublet state
would require very large basis sets of rotational functions and
could not be converged. In the present work we therefore
approximate the operator $V_{\rm int} (R,\theta)$ operator by
taking $V_S=V_{5/2}$ for all spin states. This approximation is
legitimate because we are primarily interested in N--NH
collisions between magnetically trapped atoms and molecules,
with $m_{s{\rm A}}=s_{\rm A}=\frac{3}{2}$ and $m_{s{\rm
B}}=s_{\rm B}=1$. These are {\em spin-stretched} states, and
$V_{3/2}$ and $V_{1/2}$ have no matrix elements (diagonal or
off-diagonal) involving spin-stretched states. When this
approximation is made, orthogonality relations for the $3j$
symbols reduce Eq.\ \ref{intpot} to a form diagonal both in
$m_{s{\rm A}}$ and $m_{s{\rm B}}$. The explicit expression for
$\langle n_{\rm B}  m_{n{\rm B}} L M_L  |  V_S (R,\theta) |
n_{\rm B}  m'_{n{\rm B}} L' M'_L  \rangle $ is the same as for
scattering of NH from a closed-shell atom
\cite{Gonzalez-Martinez:2007}, with the addition of factors
$\delta_{m_{s{\rm A}}{m_{s{\rm A}}'}}$.

The intermolecular spin-spin interaction has matrix elements
\begin{eqnarray}
\langle s_{\rm A} m_{s{\rm A}} s_{\rm B} m_{s{\rm B}} n_{\rm B}  m_{n{\rm B}} L M_L |V_{\rm SS}|
 s_{\rm A} m'_{s{\rm A}} s_{\rm B} m'_{s{\rm B}} n'_{\rm B}  m'_{n{\rm B}} L' M'_L \rangle =
\nonumber \\
\sqrt{30} \lambda(R) \delta_{n_{\rm B} n'_{\rm B}}
\delta_{m_{n {\rm B}} m'_{n {\rm B}}}
(-1)^{ s_{\rm A} + s_{\rm B} - m_{s{\rm A}} -  m_{s{\rm B}} -M_L }
[s_{\rm A}(s_{\rm A}+1)(2s_{\rm A} +1) s_{\rm B} (s_{\rm B} +1) (2s_{\rm B} +1)(2L+1)(2L'+1)]^{\frac{1}{2}}
\nonumber
\\
\left( \begin{array}{ccc}  L  & 2  & L' \\  0 & 0 & 0  \end{array} \right)
\sum_{q_1 q_2}
\left( \begin{array}{ccc}  L  & 2  & L' \\  -M_L & -q_1-q_2 &  M'_L  \end{array} \right)
\left( \begin{array}{ccc}  1  & 1  & 2  \\  q_1 & q_2 & -q_1-q_2  \end{array} \right) \nonumber\\
\left( \begin{array}{ccc}   s_{\rm A}   &  1   &   s_{\rm A} \\  -m_{s{\rm A}}  &  q_1   & m'_{s{\rm A}}   \end{array} \right)
\left( \begin{array}{ccc}   s_{\rm B}   &  1   &   s_{\rm B} \\  -m_{s{\rm B}}  &  q_2   & m'_{s{\rm B}}   \end{array} \right).
\end{eqnarray}
The spin-spin coupling constant $\lambda(R)$ is $E_{\rm
h}\alpha^2a_0^3/R^3$, where $E_{\rm h}$ is the Hartree energy
and $\alpha$ is the fine-structure constant.

The matrix elements for NH monomer operators are the same as
for scattering of NH from a closed-shell atom
\cite{Gonzalez-Martinez:2007}, with the addition of factors
$\delta_{m_{s{\rm A}}{m_{s{\rm A}}'}}$.

If one or both of the colliding species is not in a
spin-stretched state (with the highest possible value of
$m_S$), the system will undergo very fast spin exchange driven
by the {\em difference} between the $S=\frac{5}{2}$,
$\frac{3}{2} $ and $S=\frac{1}{2}$ potentials. For
spin-stretched states, however, spin exchange cannot occur and
only spin relaxation is possible. There are two mechanisms for
spin relaxation. The first is similar to the well-known
mechanism of spin relaxation for spin-stretched states of
alkali metal atoms, and arises through direct coupling of the
initial state $m_{s{\rm A}}=+\frac{3}{2}, m_{s{\rm B}}=+1$
(with $n_{\rm B}=m_{n{\rm B}}=0$), to final states with
$m_{s{\rm A}}$ and/or $m_{s{\rm B}}$ reduced by 1 by the
intermolecular spin-spin interaction term and $M_L$ increased
to conserve $M_{\rm tot}$. Such transitions are relatively
slow, because the intermolecular spin-spin interaction is weak.
The second mechanism is that described by Krems and Dalgarno
\cite{Krems:mfield:2004}. For the same initial state, the
intramonomer spin-spin interaction mixes $n_{\rm B}=0$ with
$n_{\rm B}=2$, and even in a magnetic field it mixes $m_{n \rm
B}=0$ with $m_{n \rm B}=\pm 1,\pm 2$. The states with $m_{n \rm
B}=+1,+2$ have $m_{s{\rm B}}=0,-1$. The states with $m_n=1,2$
are then coupled by the anisotropy of the interaction potential
to $n=0$, $m_n=0$ states with changed $M_L$ but the same
$m_{{\rm B}}$ (which is lower than in the initial state). This
mechanism is also expected to be fairly weak for a
low-anisotropy system such as N--NH: the $n=0$ and $n=2$
rotational levels of NH differ in energy by 96 cm$^{-1}$, while
the potential anisotropy $V_2(R)$ that couples them is a
short-range interaction that is never greater than 40 cm$^{-1}$
in the energetically accessible region. For both mechanisms,
spin relaxation is suppressed for s-wave scattering
($L=0,M_L=0$) at low energies and fields because the
conservation of $M_{\rm tot}$ requires $M_L'\ne0$ and therefore
$L'>0$, producing a centrifugal barrier in the outgoing
channel.

We carry out scattering calculations with the MOLSCAT package
\cite{MOLSCAT:v15}. The coupled equations are solved using the
hybrid log-derivative/Airy propagator of Alexander and
Manolopoulos \cite{Alexander:1987}. We used the fixed-step
log-derivative propagator from 2.8 \AA\ to 70 \AA\ with an
interval size of 0.08 \AA, followed by a variable-step Airy
propagation out to 400 \AA. We carried out convergence tests on
state-to-state cross sections both in the s-wave regime and at
energies up to $E=1$~K, at fields of $B=200$~G, 1000~G and 2~T.
In all cases a basis set with $n=0\ldots3$ and $L=0\ldots7$
gave convergence to within approximately 1\% for all
state-to-state cross sections. This basis set was therefore
used in all the remaining calculations.

\section{Results}

\begin{figure}
\includegraphics[width=1.1\linewidth]{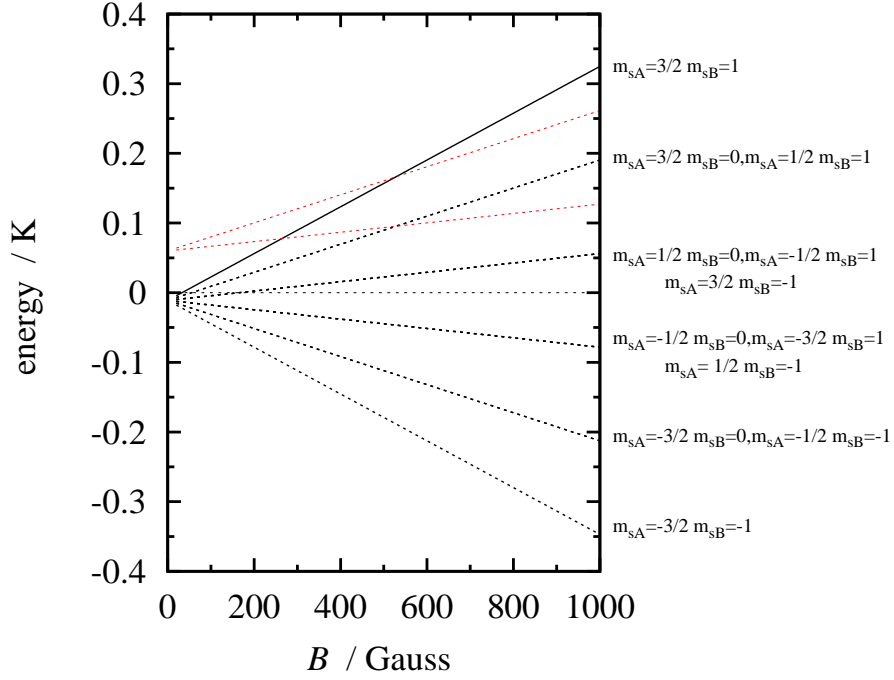}
\caption{Energy levels of the noninteracting N+NH system in a
magnetic field. The dotted red lines show the energy obtained by
adding the d-wave centrifugal barrier height (71~mK) to the
levels with $M_{\rm tot}=\frac{3}{2}$ and $M_{\rm tot}=\frac{1}{2}$. The
crossings between the red lines and the initial-state energy
indicate the fields above which s-wave inelastic cross sections
are no longer suppressed by centrifugal barriers.
} \label{thres}
\end{figure}

Fig.\ \ref{thres} shows the Zeeman energy levels of the
noninteracting N+NH system. In the buffer-gas cooling
experiment, \cite{Hummon:2009} both atoms and molecules are
trapped in their low-field-seeking state with $m_{s{\rm
A}}=\frac{3}{2}$ and $n_{\rm B}=m_{n{\rm B} }=0$, $m_{s{\rm
B}}=1$. The experiment has already achieved temperatures around
550~mK, and at this temperature atoms with an energy of $5kT$
sample magnetic fields up to 2~T in a quadrupole trap. However,
as the temperature decreases, so too will the magnetic fields
sampled. We therefore consider collision energies from
10~$\mu$K to 1~K and fields from 10~G to 2~T.

The N atom is considerably less polarizable than alkali metal
or alkaline earth atoms. As a result, the dispersion
coefficient $C_{6,0}$ for N--NH is considerably lower than for
most metal atom -- molecule systems that have been considered
previously as candidates for sympathetic cooling. Together with
a low reduced mass, this results in relatively high centrifugal
barriers for $L>0$ partial waves: 14 mK for $L=1$, 71 mK for
$L=2$, 120 mK for $L=3$, etc. The high centrifugal barriers
also mean that quite small number of partial waves are needed
to converge cross sections: for example, including
contributions from $L$ up to 4 is sufficient to obtain
convergence up to about 0.5~K.

\subsection{ Close-coupling  calculations }

\begin{figure}
\includegraphics[width=0.5\linewidth]{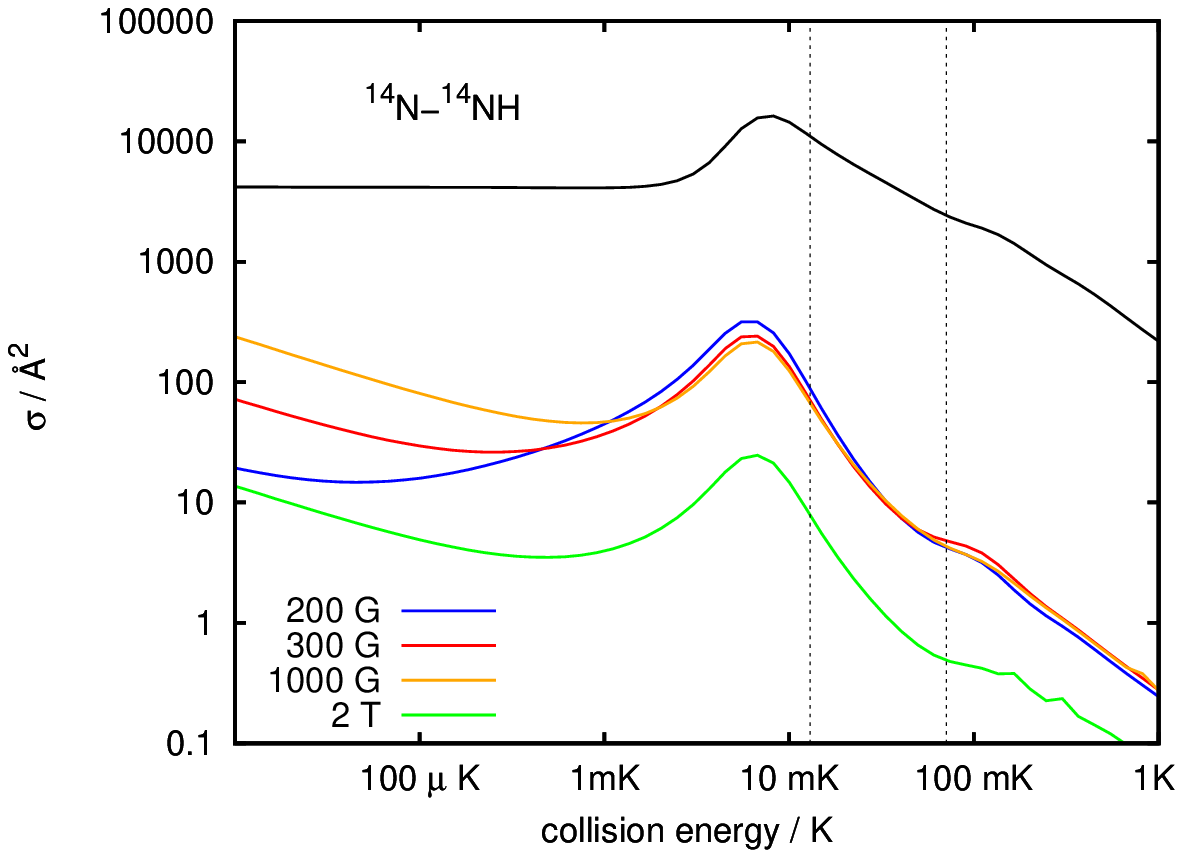}
\includegraphics[width=0.5\linewidth]{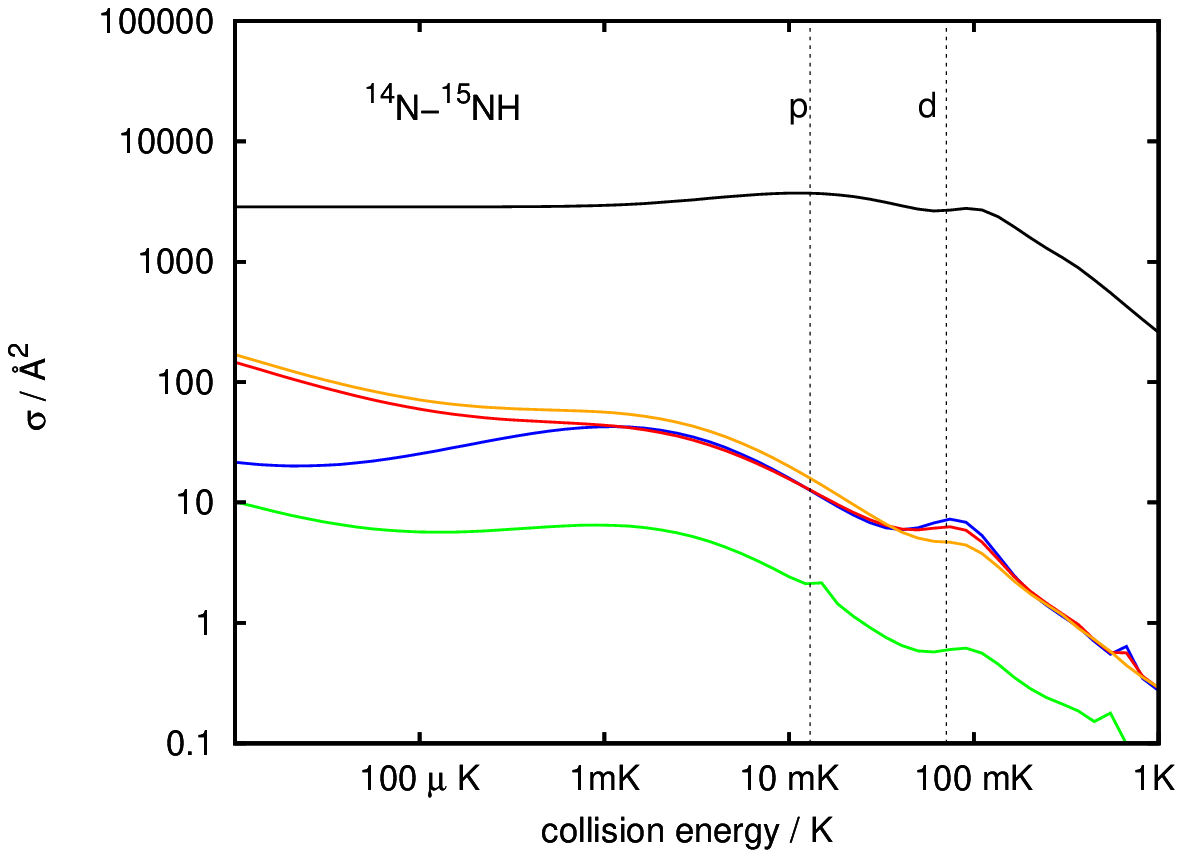}
\includegraphics[width=0.5\linewidth]{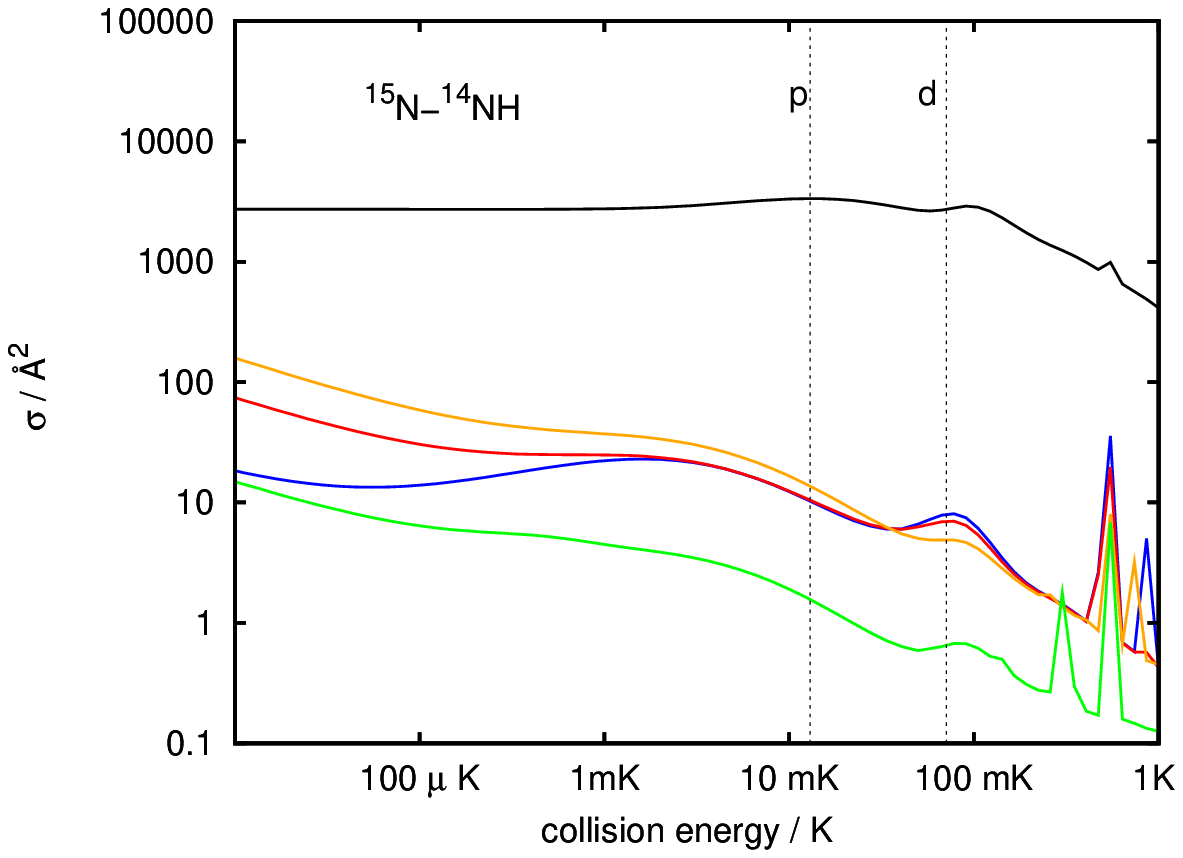}
\includegraphics[width=0.5\linewidth]{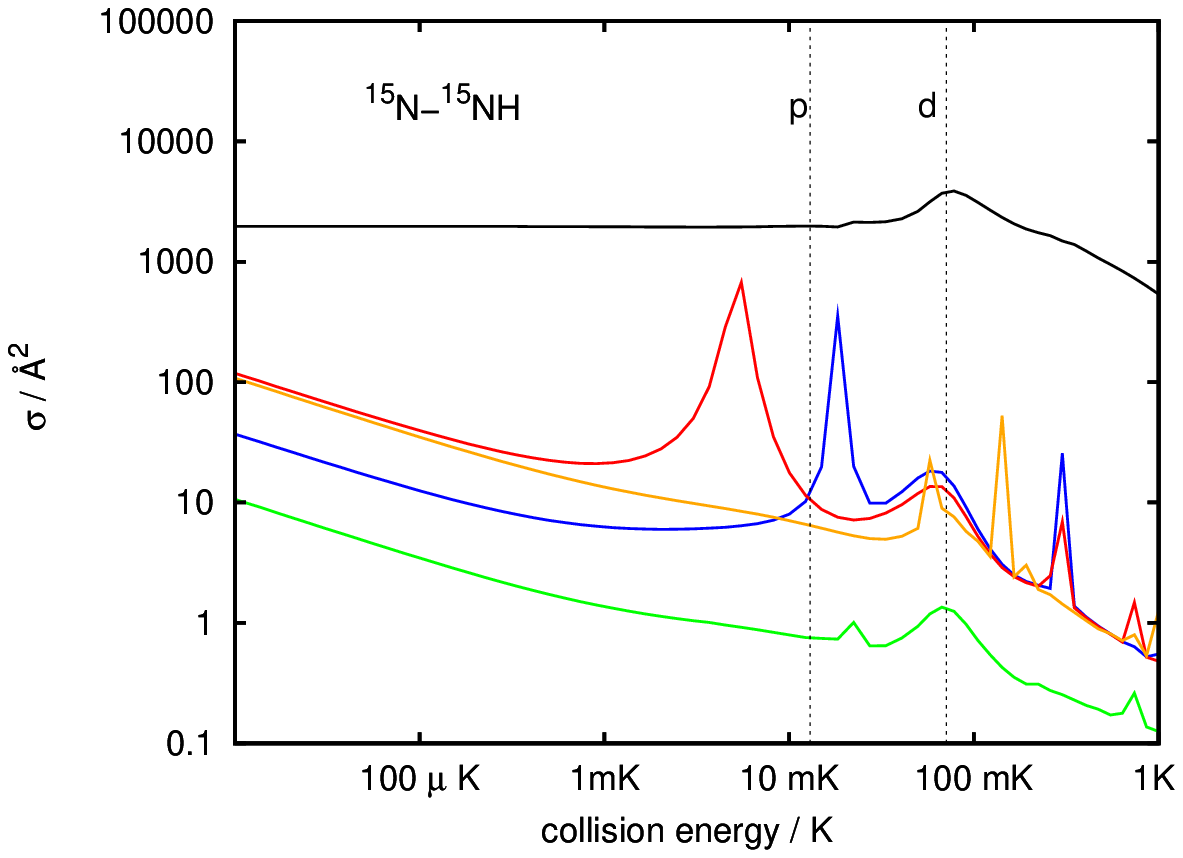}
\caption{Elastic and inelastic cross sections for N--NH
scattering for $^{14}$N--$^{14}$NH, $^{14}$N--$^{15}$NH,
$^{15}$N--$^{14}$NH and $^{15}$N--$^{15}$NH for different
magnetic fields. The elastic cross section (black line) is
almost independent of field. The positions of the p- and d-wave
exit-channel barriers are marked with vertical lines.}
\label{nnhxs}
\end{figure}

\begin{figure}
\includegraphics[width=0.5\linewidth]{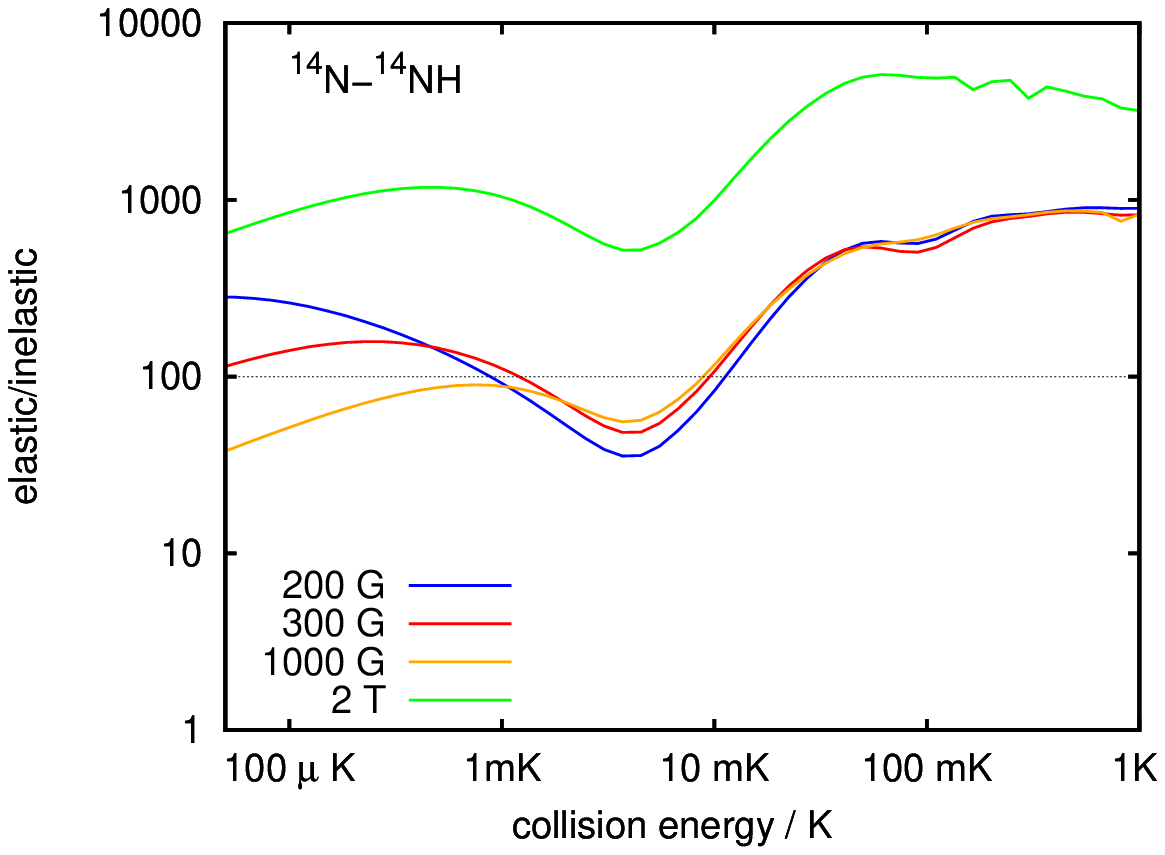}
\includegraphics[width=0.5\linewidth]{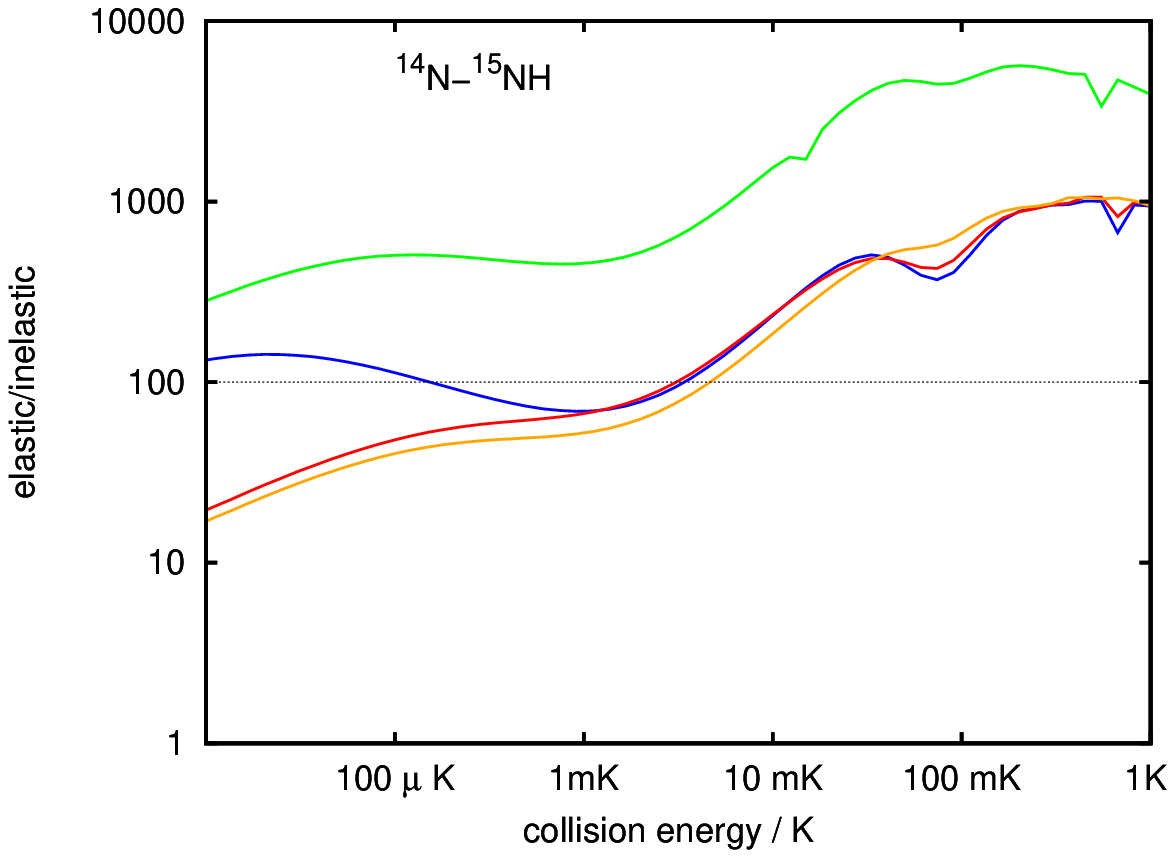}
\includegraphics[width=0.5\linewidth]{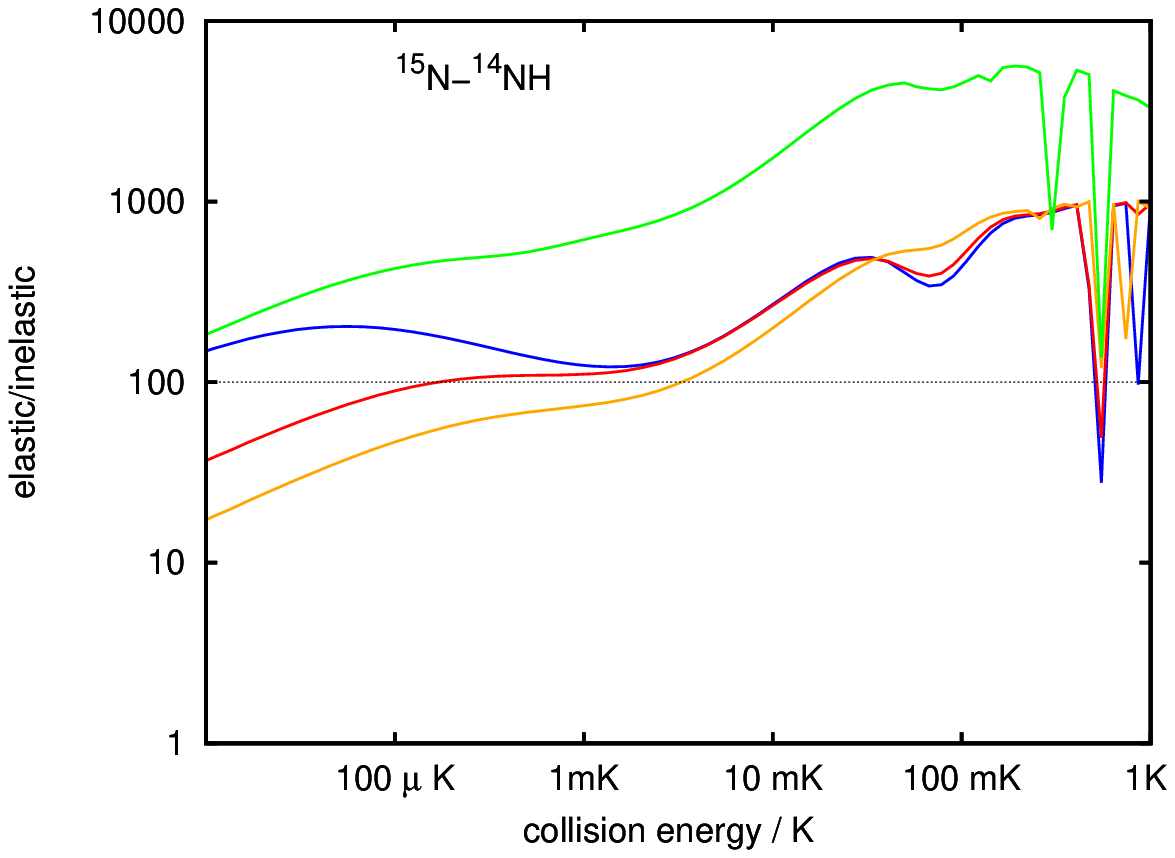}
\includegraphics[width=0.5\linewidth]{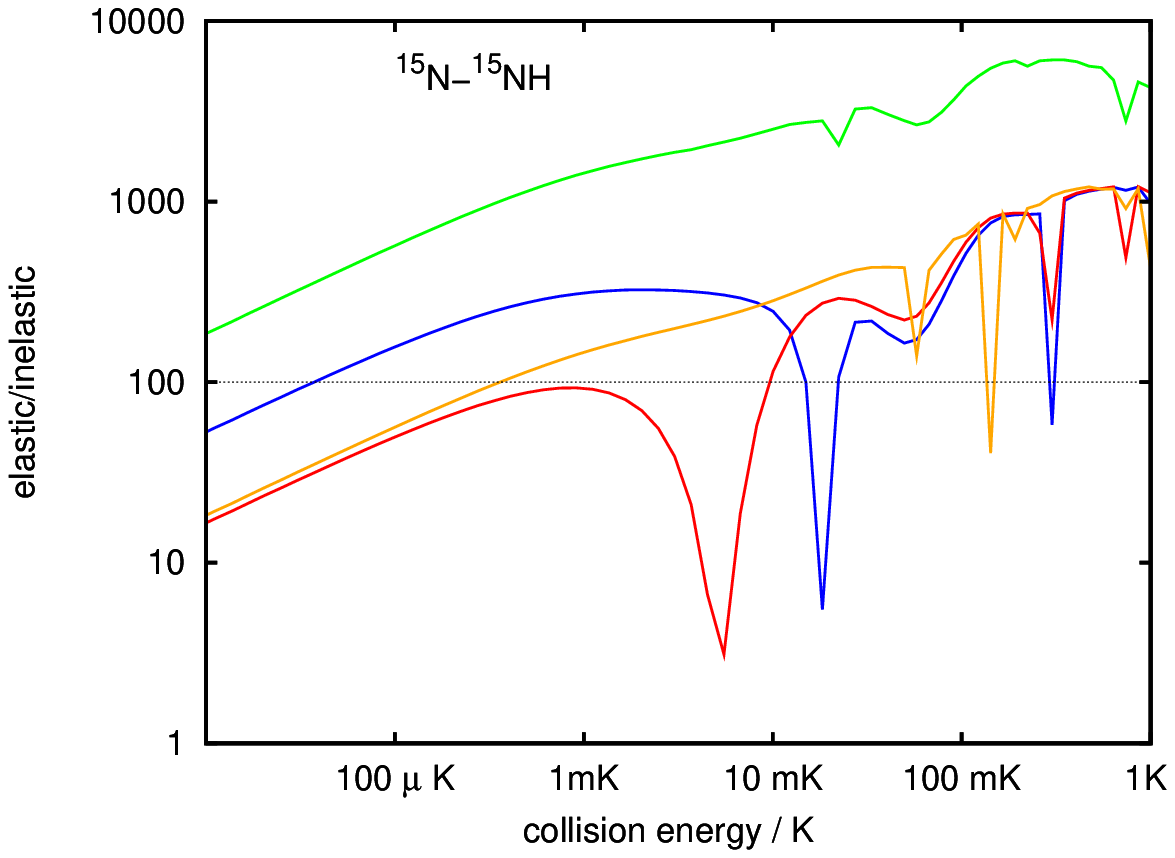}
\caption{The ratio of elastic to inelastic cross sections for
$^{14}$N~--~$^{14}$NH, $^{14}$N~--~$^{15}$NH,
$^{15}$N~--~$^{14}$NH and $^{15}$N~--~$^{15}$NH systems for
different magnetic fields. The value of 100 typically required
for sympathetic cooling is marked with a dotted horizontal
line.} \label{nnhgamma}
\end{figure}

\begin{figure}
\includegraphics[width=1.2\linewidth]{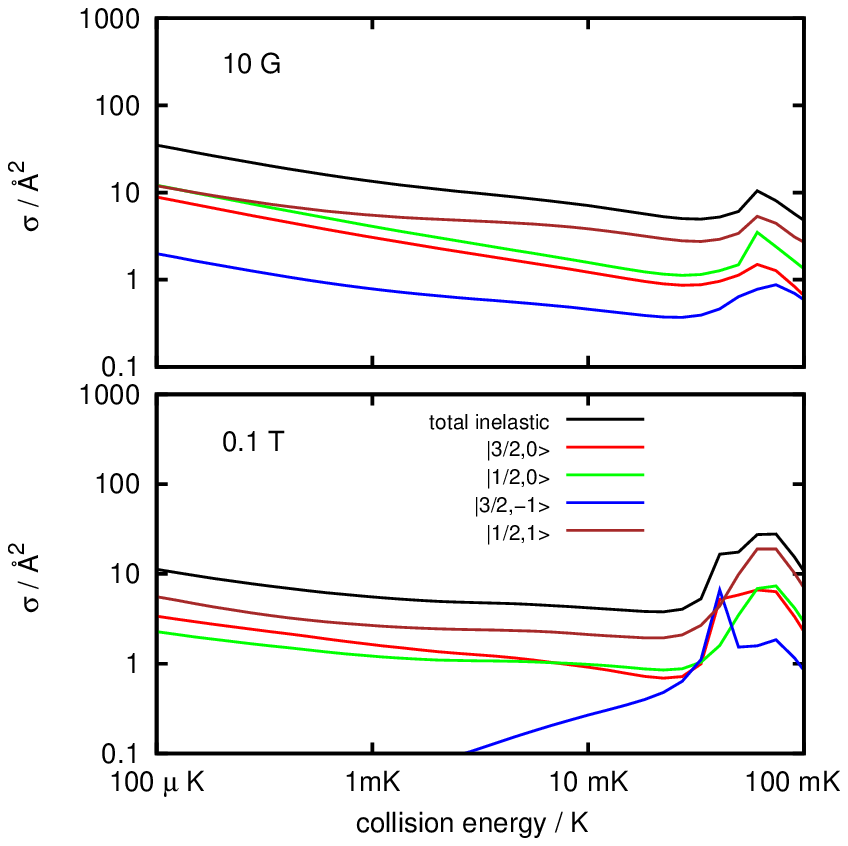}
\caption{State-to-state inelastic cross sections for $^{15}$N~--~$^{15}$NH
for weak (10~G) and strong (0.1~T) magnetic fields.}
\label{nnhstate-to-state}
\end{figure}

Calculated elastic and inelastic cross sections for the four
different isotopic combinations are shown as a function of
collision energy $E$ in Fig.\ \ref{nnhxs}, for representative
magnetic fields of 200~G, 300~G, 1000~G and 2~T. In a simple
hard-sphere model of sympathetic cooling, neglecting inelastic
collisions, the temperature relaxes towards equilibrium and
reaches a 1/$e$ point after $(m_1+m_2)^2/2m_1m_2$ collisions
\cite{deCarvalho:1999}, where $m_1$ and $m_2$ are the masses of
the two species. For sympathetic cooling to be successful we
need the ratio of elastic to inelastic cross sections to be
much larger than this. The calculated ratios are shown in Fig.\
\ref{nnhgamma}: for the most part they are more than 50 at
collision energies above about 1 mK, indicating that
sympathetic cooling of NH by N is likely to be feasible.

Several different effects are evident in Fig.\ \ref{nnhxs}. The
first is that the cross sections enter the s-wave regime, where
they are proportional to $E^{-1/2}$, at quite different
energies for different isotopic species. This occurs because of
p-wave resonant effects. Once in the s-wave regime, however,
the inelastic cross sections generally decrease at magnetic
fields below about 300~G, because of the centrifugal barriers
in the outgoing channels. Since atoms and molecules in a
quadrupole trap sample lower and lower fields as the
temperature is decreased, this indicates that sympathetic
cooling will become increasingly effective as the temperature
is lowered, as predicted for Mg-NH \cite{Wallis:MgNH:2009}.
Lastly, inelastic collisions are also suppressed for very {\em
high} magnetic fields. All these effects will be discussed in
more detail below.

Spin relaxation collisions can change $m_{s{\rm A}}$ for the N
atom, $m_{s{\rm B}}$ for the NH molecule, or both. Fig.\
\ref{nnhstate-to-state} shows the state-to-state cross sections
for the most important final states for $^{15}$N--$^{15}$NH as
a function of energy at two different fields. It may be seen
that dominant final states are those in which $m_{s{\rm A}}$
and/or $m_{s{\rm B}}$ has changed by 1. These collisions are
driven by the intermolecular spin-spin interaction. Transitions
that change $m_{s{\rm B}}$ by 2 can occur only by the second
mechanism described in Section \ref{sec:theory} above,
involving the potential anisotropy, and are seen to be very
much weaker except in a small resonant region.

The $^{14}$N--$^{14}$NH system shows behaviour quite different
from the others, with a large peak in the inelastic cross
sections near 10~mK which reduces the elastic-to-inelastic
ratio to around 10. This ratio may not be high enough for
effective sympathetic cooling from an initial temperature of
tens of milliKelvin. The peak appears at the same energy for
all values of the field. It arises from a p-wave shape
resonance in the incoming channel, as discussed in section
\ref{sec:aqdt} below. For the larger reduced masses of the
other isotopic combinations, the quasibound state responsible
for the shape resonance drops below threshold and becomes a
true bound state. Thus the other isotopic combinations do not
exhibit this feature and have more favourable properties for
sympathetic cooling.

The $^{15}$N--$^{15}$NH system exhibits d-wave shape resonances
for collision energies of 50 to 70 mK, but they are much weaker
than the p-wave resonance for $^{14}$N--$^{14}$NH and their
presence does not strongly affect the total inelastic cross
section.

\begin{figure}
\includegraphics[width=0.85\linewidth]{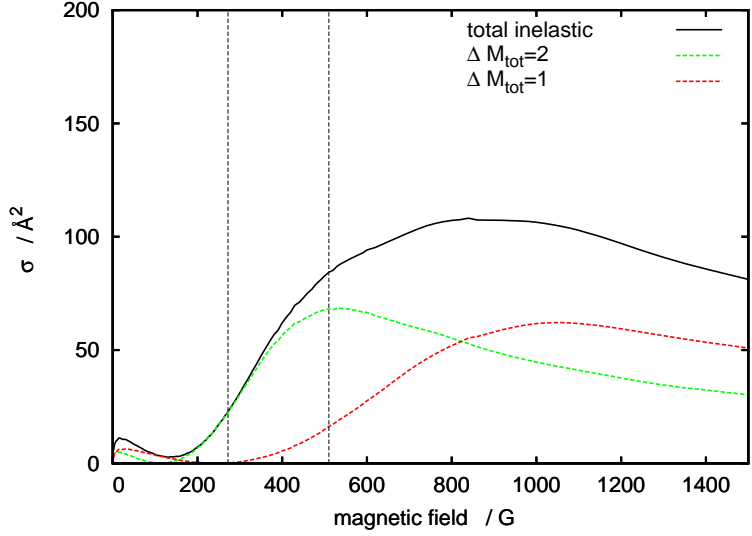}
\caption{Inelastic s-wave cross section as a function of
magnetic field for $^{14}$N--$^{14}$NH at a collision energy of
50~$\mu$K. The dashed vertical lines correspond to the fields for
which the kinetic energy release from $M_{\rm tot}=\frac{5}{2}$ to
$M_{\rm tot}=\frac{3}{2}$ and $M_{\rm tot}=\frac{1}{2}$ states are equal to the
height of the d-wave barrier.} \label{swfld}
\end{figure}

The $L=2$ centrifugal barrier plays a crucial role in spin
relaxation in the ultracold regime. For an incoming channel
with $L=0$, spin relaxation requires outgoing $L\ge2$. If the
energy difference between the incoming and outgoing channels is
smaller than the height of the $L=2$ centrifugal barrier, the
s-wave inelastic cross section is suppressed (see Fig.\
\ref{thres}). The s-wave state-to-state cross sections are
shown as a function of magnetic field in Fig.\ \ref{swfld} for
$^{14}$N--$^{14}$NH at a collision energy of 50~$\mu$K. The
inelastic cross sections generally decrease at magnetic fields
below about 500~G, though there is a dip in each state-to-state
component between 100 and 300~G. These dips are due to
suppression of the inelastic cross sections in the wings of
resonances, as described by Hutson {\em et al.}\
\cite{Hutson:HeO2:2009}; in this case the resonances concerned
are shape resonances in the d-wave outgoing channels.

The suppression of inelastic scattering by the centrifugal
barrier is clear in the total inelastic cross section only for
systems with reduced mass larger than for $^{14}$N--$^{14}$NH.
For the $^{14}$N--$^{14}$NH system itself, the p-wave
contribution is very strong even at very low energies. In fact,
the p-wave enhancement of the inelastic cross section between
50 $\mu$K and 1~mK is so strong that the total inelastic cross
section does not follow the $E^{-\frac{1}{2}}$ power-law
dependence expected from the Wigner threshold laws at these
energies.

Suppression of inelastic collisions due to barriers in the
outgoing channels decreases as the magnetic field increases (so
that the kinetic energy release increases). Eventually,
however, the inelastic cross section reaches a maximum and
starts to decrease again. This occurs for all partial waves,
and the total cross sections at a field of 2~T are typically
reduced by a factor of about 10 from their values at 0.1~T.
Since for some isotopic combinations the low-field ratio of
elastic to inelastic cross sections may not be large enough at
temperatures of 1 to 10~mK, the application of a strong bias
field to a magnetic trap offers a possible way way to improve
the ratio.

\begin{figure}
\includegraphics[width=0.85\linewidth]{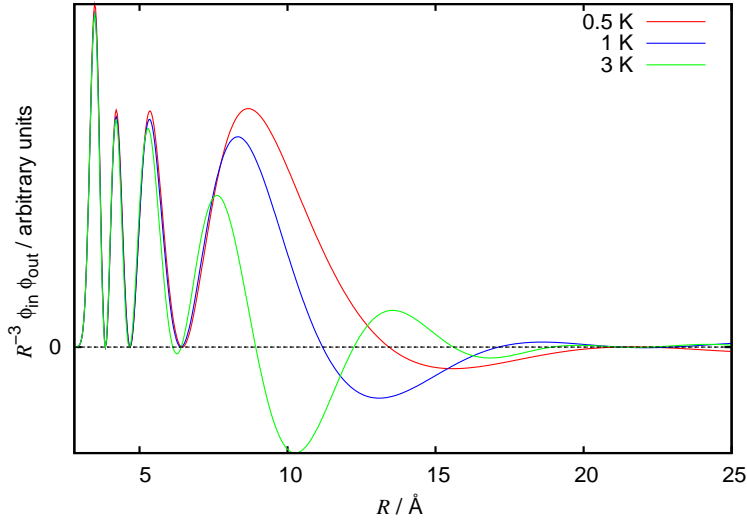}
\caption{The integrand of the distorted-wave Born approximation
(Eq.\ \ref{sif}) for $^{14}$N--$^{14}$NH at a collision energy of
1~mK for different kinetic energy releases. The incoming wavefunction
is calculated for $L=0$ and the outgoing wavefunction for $L=2$.}
\label{Born}
\end{figure}

Suppression of inelastic cross sections at high fields has been
observed for O($^3$P)-He collisions \cite{Krems:OHe:2002}, for
OH-OH \cite{Ticknor:OHmag:2005} and for collisions of Cr atoms
\cite{Pavlovic:2005}. For small inelasticity, the
distorted-wave Born approximation gives \cite{Child:1974}
\begin{equation}
\sigma_{i\to f} = 4\pi k_i^{-2}
\left| \int \psi_i(R) U_{if}(R) \psi_f(R)\,dR  \right|^2,
\label{sif}
\end{equation}
where $\psi_i$ and $\psi_f$ are energy-normalized wavefunctions
in the initial and final channels, $U_{if}$ is the coupling
between the channels, and $k_i$ is the wave vector in the
incoming channel. Fig.\ \ref{Born} shows the integrand of Eq.\
\ref{sif} for the intermolecular spin-spin term ($R^{-3}$) at
kinetic energy releases of 0.5~K, 1~K and 3~K, corresponding to
fields of 3800~G, 7600~G and 2.3~T for transitions with $\Delta
m_{s{\rm A}}+\Delta m_{s{\rm B}}=-1$. It may be seen that there
is significant oscillatory cancellation in the integral at high
fields, when $\psi_i(R)$ and $\psi_f(R)$ oscillate out of phase
with one another in the interaction region, and this combines
with the effect of the resonances in the d-wave outgoing
channels to produce the maxima in Fig.\ \ref{swfld}. The
oscillatory cancellation occurs for arbitrary partial waves.

\begin{figure}
\includegraphics[width=0.90\linewidth]{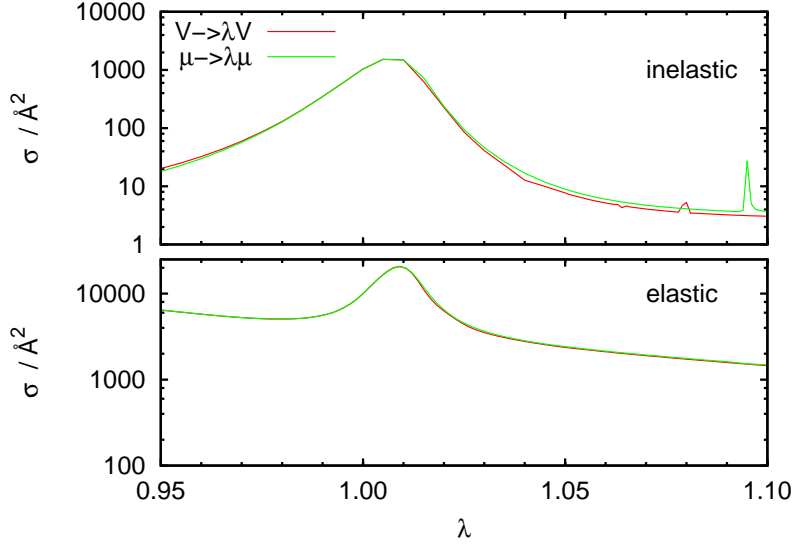}
\caption{Elastic and total inelastic cross sections for
$^{14}$N--$^{14}$NH at $E=5$~mK and $B=50$~G as a function of
both reduced mass and potential scaling factors.} \label{scalings}
\end{figure}

\subsection{Dependence on interaction potential and reduced
mass}

Because of the crucial role played by the p-wave shape
resonance for $^{14}$N--$^{14}$NH at energies up to 10 mK, it
is important to investigate the influence of uncertainties in
the interaction potential on the cross sections. The shape of
the 2D interaction potential is complicated and the scattering
properties might in principle depend on many parameters.
However, Gribakin and Flambaum \cite{Gribakin:1993} showed that
for single-channel scattering the scattering length $a$ behaves
as
\begin{equation}
 a = \bar{a} \left[ 1 -
 \mbox{tan} \left( \Phi -\frac{\pi}{8} \right) \right],
\end{equation}
where, for a potential with long-range form $-C_6R^{-6}$, the
mean scattering length $\bar{a}$ is $0.956 (2\mu
C_6/\hbar^2)^{\frac{1}{4}} $ and
\begin{equation}
\Phi =\int_0^\infty
\left( 2 \mu V_{\rm int}(R) /\hbar^2 \right)^{\frac{1}{2}}\, dR.
\label{Phieq}
\end{equation}
Although N--NH is a many-channel scattering problem, it is
elastically dominated and Eq.\ \ref{Phieq} with $V_{\rm
int}(R)$ replaced by $V_0(R)$ reproduces the major features of
the elastic scattering. Thus scaling $\mu$ is approximately
equivalent to scaling the entire interaction potential, and
either of these scalings provides a good way to explore the
variation of scattering length as a function of potential.
Fig.\ \ref{scalings} shows the elastic and total inelastic
cross sections for $^{14}$N--$^{14}$NH at $E=5$~mK and $B=50$~G
as a function of both reduced mass and potential scaling
factors. It may be seen that the two scalings have a very
similar effect, apart from a small shift in the Feshbach
resonance around $\lambda=1.08$, which comes from a
rotationally excited state of NH.

\begin{figure}
\includegraphics[width=\linewidth]{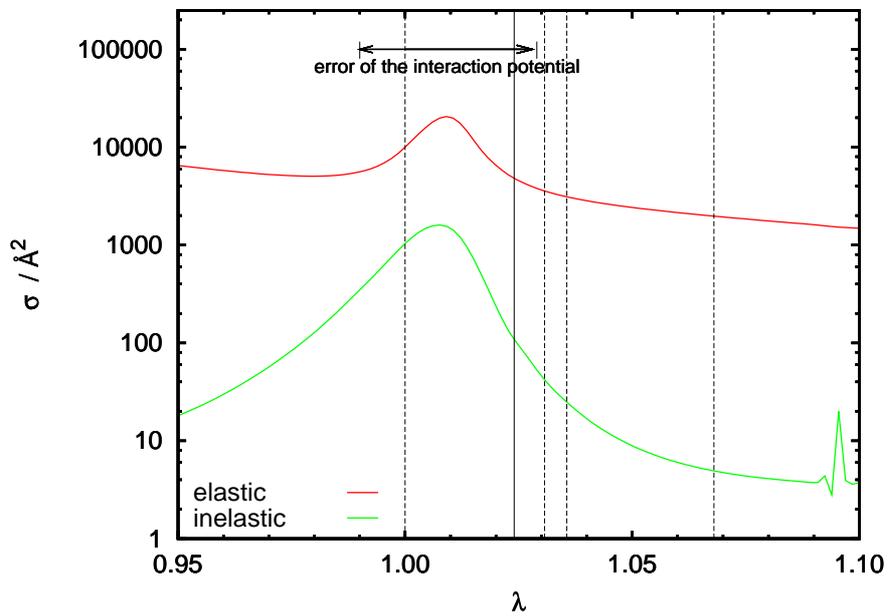}
\caption{Elastic and inelastic cross sections for a collision
energy of 5 mK as a function of scaled reduced mass. The dashed
vertical lines indicate the values of $\mu$ for the systems
concerned here, from left to right, $^{14}$N--$^{14}$NH,
$^{14}$N--$^{15}$NH, $^{15}$N--$^{14}$NH and
$^{15}$N--$^{15}$NH, respectively. The solid vertical line
shows the value $\lambda = 1.024$ for which the scattering
length $a=2\bar{a}$. } \label{xs5mK}
\end{figure}

We estimate the bounds on the accuracy of our potential to be
between $-1$ and +3\% of the well depth. We have therefore
carried out calculations of the cross sections as a function of
a variable reduced mass, parameterized by scaling factor $\mu
\to \lambda \mu$ for a collision energy of 5~mK. A scaling
factor $\lambda=1$ corresponds to the reduced mass for
$^{14}$N--$^{14}$NH. The result is shown in Fig.\ \ref{xs5mK}.
There is a strong maximum in the total inelastic cross section
near $\lambda=1.012$, due to a p-wave shape resonance in the
{\em elastic} channel. A change of 1.2\% in the potential is
within the estimated error bound of our calculations.
Enhancement of the cross sections due to the p-wave resonance
might thus occur for heavier isotopic combinations than
$^{14}$N--$^{14}$NH if our potential is slightly too deep.
Although we believe it is more likely that our potential is too
shallow than too deep, this cannot be ruled out. However, it is
quite unlikely that enhancement due to the p-wave resonance
would occur for the heaviest system, $^{15}$N--$^{15}$NH.

\subsection {AQDT analysis of shape resonances \label{sec:aqdt}}

\begin{figure}
\includegraphics[width=0.85\linewidth]{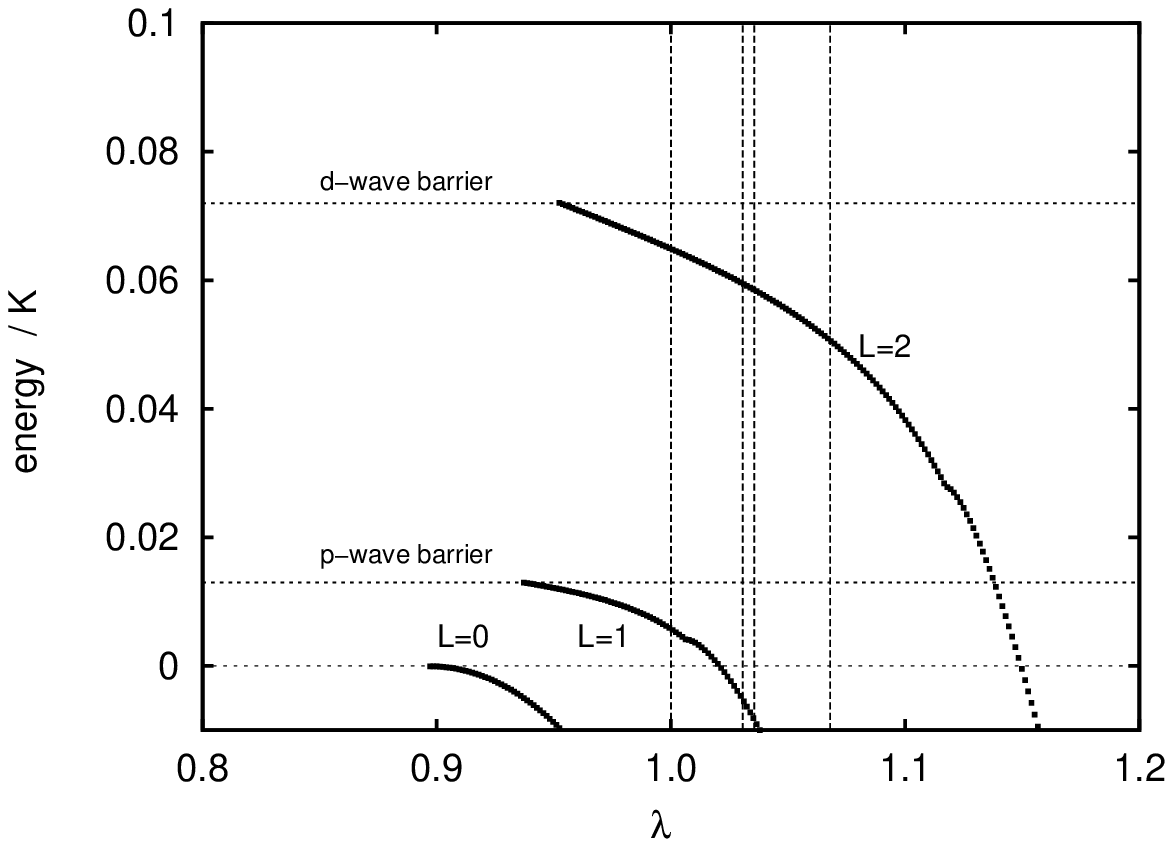}
\includegraphics[width=0.85\linewidth]{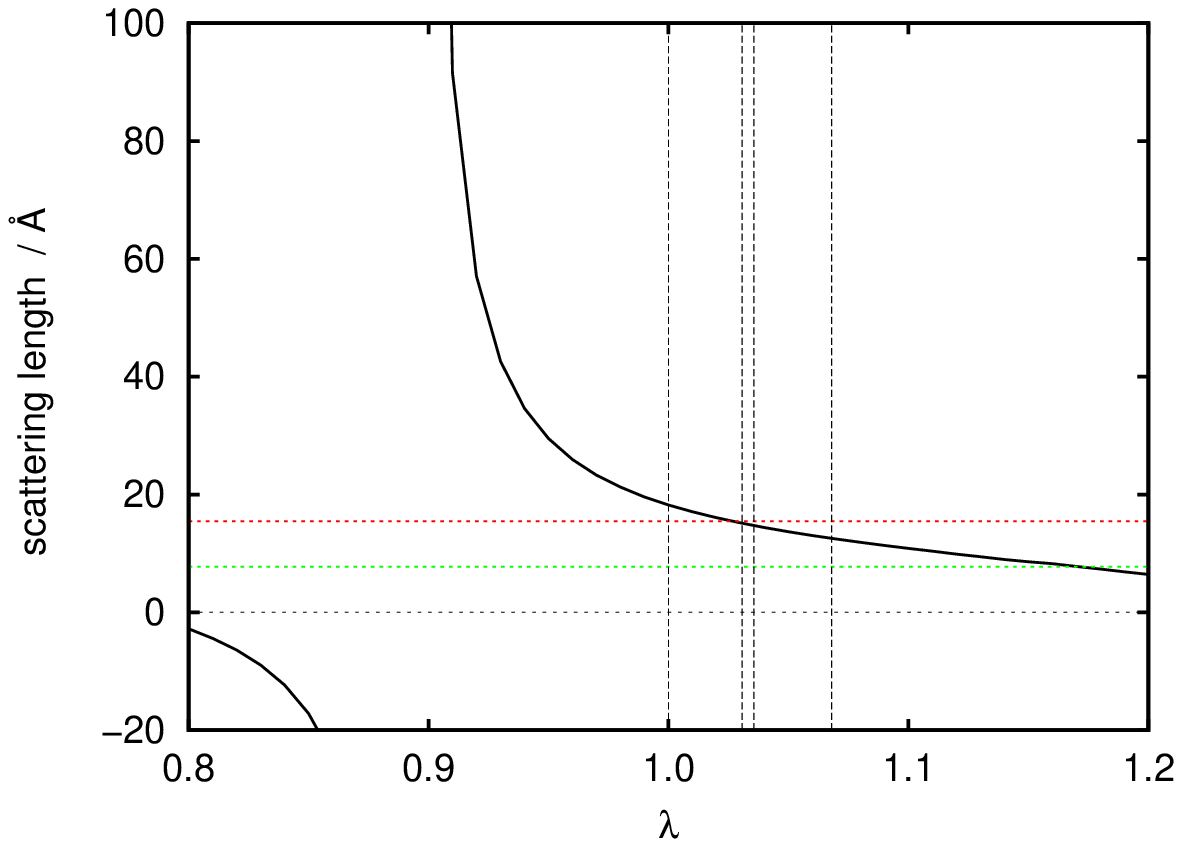}
\caption{Near-threshold bound states of sextet N--NH for
$L=0\ldots2$ as a function of $\mu$ (upper panel), and the
$\mu$-dependence of the s-wave scattering length (lower panel).
Field-free calculations with a structureless atom and diatom
were used to obtain the bound states here. The dashed vertical
lines indicate the values of $\mu$ for $^{14}$N--$^{14}$NH,
$^{14}$N--$^{15}$NH, $^{15}$N--$^{14}$NH and
$^{15}$N--$^{15}$NH, respectively, from left to right. The red
dotted line on the lower panel indicates $2\bar{a}$ and the
green dotted line indicates $\bar{a}$. } \label{nnhbnd}
\end{figure}

In this section we consider the N--NH scattering in the context
of angular-momentum-insensitive quantum defect theory (AQDT)
\cite{Gao:2001}. The upper part of Fig.\ \ref{nnhbnd} shows the
positions of (quasi)bound states for $L=0\ldots2$ as a function
of the reduced-mass scaling factor $\lambda$ for values between
0.8 and 1.2. The $L=0$ bound state crosses the threshold at a
value of the reduced mass much smaller than that for
$^{14}$N--$^{14}$NH, well outside the estimated error bounds
for the potential. There is thus no s-wave resonance in the
scattering for any of the systems considered here.

The p-wave shape resonance in the cross sections for
$^{14}$N--$^{14}$NH arises from the quasibound state with
$L=1$, which is 6 mK above threshold for $\lambda=1$. As the
reduced mass increases above this, the $L=1$ bound state
crosses the threshold at $\lambda = 1.024$ and becomes a "real"
bound state. This explains why we see no p-wave shape
resonances in the cross sections for collisions with reduced
mass larger than $1.024 \mu_{14,14}$, i.e. for all systems
containing at least one $^{15}$N atom. For large values of
$\mu$ the $L=2$ bound state comes close to threshold, and this
results in the (small) d-wave shape resonance that can be seen
in the $^{15}$N--$^{15}$NH cross sections.

The lower panel of Fig.\ \ref{nnhbnd} shows the s-wave
scattering length as a function of scaling factor $\lambda$. As
expected, there is a pole near $\lambda=0.9$ as the $L=0$ bound
state crosses the threshold. Within the estimated error bounds
of the potential, $a$ varies between 22.3 and 15.2 \AA, so that
the elastic cross section for small $E_{\rm coll}$ (in the
Wigner regime) varies between 6250 and 2900 \AA$^2$.

In angular-momentum-insensitive quantum defect theory
\cite{Gao:1998, Gao:2000, Gao:2001}, the scattering properties
of a system for arbitrary $L$ can be predicted from only a few
parameters: the s-wave scattering length $a$, the dispersion
coefficient $C_6$ and the reduced mass $\mu$ (and thus
$\bar{a}$). The positions where $L>0$ bound states cross
threshold, and hence produce shape resonances, depend only on
the relationship between $a$ and $\bar{a}$. In particular, when
$a=2 \bar{a}$ there is an $L=1$ bound state exactly at
threshold, and systems with $a$ slightly larger than $2
\bar{a}$ have a p-wave shape resonance at a collision energies
below the height of the p-wave centrifugal barrier. This is the
case for the $^{14}$N--$^{14}$NH system here.  For $a=\bar{a}$
there is an $L=2$ bound state exactly at threshold. The
scattering length for $^{15}$N--$^{15}$NH is $1.54 \bar{a}$,
which is close enough above $\bar{a}$ to produce a d-wave shape
resonance at finite energy. The energies at which the p-wave
and d-wave resonances appear can be read off the $L=1$ and 2
lines in Fig.\ \ref{nnhbnd}.

It should be noted that a change in the interaction potential
would result in ``sliding" the vertical lines representing the
four isotopic combinations horizontally along Fig.\
\ref{nnhbnd}. A range of behaviour can exist for potentials
within our uncertainties of our calculations. Our calculations
thus do not definitively identify which characteristics will be
be observed for a particular isotopic combination.

\begin{figure}
\includegraphics[width=0.85\linewidth]{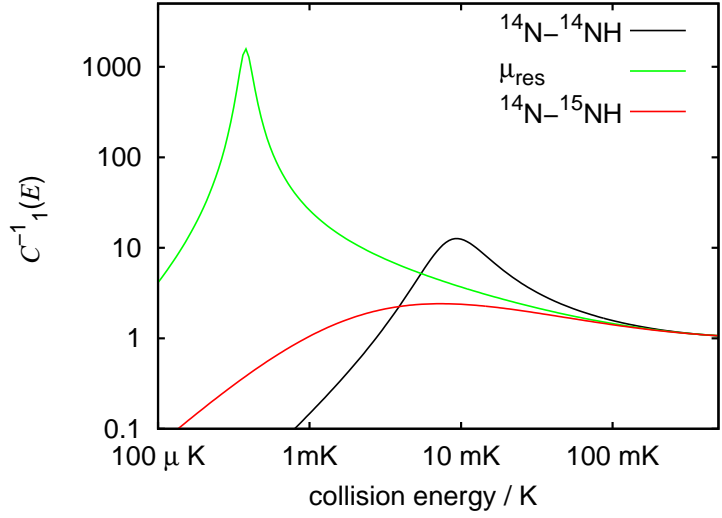}
\caption{The p-wave transmission function $C_1^{-1}(E)$ for
N--NH systems with reduced masses corresponding to:
$^{14}$N--$^{14}$NH ($a/\bar{a}=2.34$), $^{14}$N--$^{15}$NH
($a/ \bar{a}=1.94$), and an artificial system with
$\lambda=1.031$, for which $a/\bar{a}\to 2^+$. } \label{nnhc1}
\end{figure}

\begin{figure}
\includegraphics[width=0.85\linewidth]{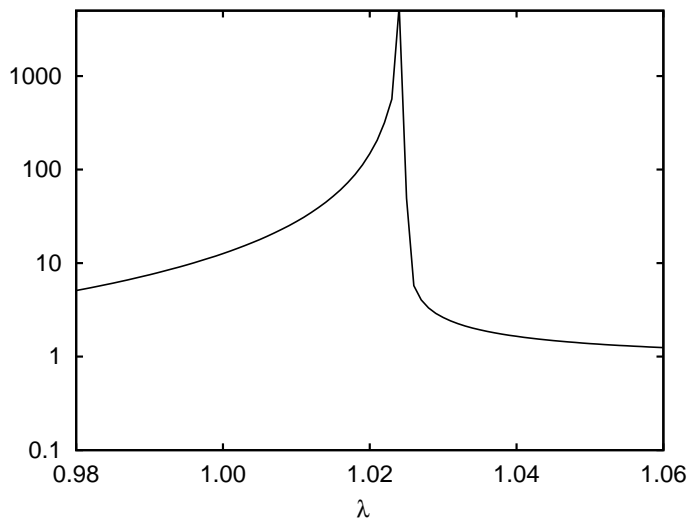}
\caption{The height of the peak in the $C_1^{-1}(E)$ function
as a function of reduced-mass scaling factor $\lambda$. }
\label{nnhc1max}
\end{figure}

To study the shape of the resonant features more
quantitatively, we can consider the $C_L(E)$ functions
introduced by Mies \cite{Mies:1984a, Mies:1984}. These
functions give the connection between a semiclassical JWKB
description of scattering states (valid at large collision
energies) and the near-threshold behaviour. The function
$C_L^{-1}(E)$ can be viewed as an enhancement factor in the
short-range part of the wavefunction due to the presence of the
long-range potential, including any resonant effects in the
incoming channel. In Fig.\ \ref{nnhc1} we show the
$C_1^{-1}(E)$ functions for p-wave scattering with $\lambda=1$,
$1.023$ and 1.031 (the last of these values corresponding to
the reduced mass of $^{14}$N--$^{15}$NH). As the scattering
length decreases and reaches $2\bar{a}$, the height of the peak
in $C_1^{-1}(E)$ goes to $+\infty$, as shown in Fig.\
\ref{nnhc1max}, and the energy at which the peak occurs
approaches zero. The width of the resonance decreases,
corresponding to increasing the lifetime of the quasibound
state. The intensity of the resonance rapidly decreases once
the scattering length is larger than $2\bar{a}$, corresponding
to a bound state below threshold.

\section{Conclusions}
We have calculated a potential energy surface for N atoms
($^4S$) interacting with NH molecules ($^3\Sigma^-$) in the
spin-$\frac{5}{2}$ (sextet) state, using unrestricted
coupled-cluster calculations with an explicitly correlated
basis set. This is the surface that governs collisions of cold
N atoms and NH molecules in a magnetic trap. We have used the
surface to carry out quantum scattering calculations of cold
collisions for different isotopic combinations of N and NH, as
a function of collision energy and magnetic field.

The sextet potential energy surface is weakly anisotropic, with
an anisotropy of approximately 40 cm$^{-1}$ in the well region.
The anisotropy is dominated by the $P_2(\cos\theta)$ Legendre,
which mixes states with $\Delta n=\pm2$ in the NH rotational
quantum $n$. Since the anisotropy is smaller than the
separation between the $n=0$ and 2 states, it causes relatively
weak mixing during collisions and the scattering is generally
elastically dominated. The inelastic cross sections are
suppressed both at low energy and low field (by centrifugal
barrier in the exit channels) and at very high field (by
oscillatory cancellation due to the large kinetic energy
release). For most isotopic combinations the ratio of elastic
to inelastic cross sections is high enough, over a wide enough
range of collision energy and magnetic field, that sympathetic
cooling of NH by N is a good prospect.

We have shown that a scaling of the interaction potential is
approximately equivalent in its effects to a scaling of the
collision reduced mass. We estimate our interaction potential
to be accurate to within 3\%. We have investigated scaling the
reduced mass by up to 20\% from the value for
$^{14}$N--$^{14}$NH. The scaling revealed that there are major
effects arising for a p-wave shape resonance, which produce
enhanced inelastic scattering for $^{14}$N--$^{14}$NH at low
energies on our best potential surface. We have used
angular-momentum-insensitive quantum defect theory (AQDT) to
understand how the results change for different isotopic
combinations.

Scaling the potential energy surface, or equivalently the
reduced mass, is a very useful tool for understanding cold
collision calculations. In combination with AQDT, it can
provide powerful insights into low-energy scattering for
low-energy collisions where where only a few partial waves
contribute to the scattering.

\section*{Acknowledgments}

We are grateful to John Doyle for interesting us in this
problem, and to Alisdair Wallis for valuable discussions. This
work is supported by EPSRC under collaborative project CoPoMol
of the ESF EUROCORES Programme EuroQUAM.


\bibliographystyle{rsc}
\providecommand*{\mcitethebibliography}{\thebibliography}
\csname @ifundefined\endcsname{endmcitethebibliography}
{\let\endmcitethebibliography\endthebibliography}{}

\end{document}